\documentclass[preprint,12pt]{elsarticle}

\usepackage{amssymb}

\journal{Physics Letters B}

\begin{document}

\begin{frontmatter}



\title{Quasifission and difference in formation of evaporation residues
in the $^{16}$O+$^{184}$W and $^{19}$F+$^{181}$Ta reactions}


\author[a1,a2]{A. K. Nasirov}
\ead{nasirov@jinr.ru}
\address[a1]{Joint Institute for Nuclear Research, 141980 Dubna, Russia}
\address[a2]{Institute of Nuclear Physics, 100214, Tashkent, Uzbekistan}
\author[a3]{G. Mandaglio}
\author[a3]{M. Manganaro}
\author[a2]{A. I. Muminov}
\author[a3]{G. Fazio}
\address[a3]{Dipartimento di Fisica dell' Universit\`a di Messina, 98166 Messina,  and Istituto Nazionale di Fisica Nucleare, Sezione di Catania,  Italy}
\author[a3]{G. Giardina}
\ead{giardina@nucleo.unime.it}

\begin{abstract}
The  excitation functions of capture, complete fusion, and evaporation residue formation in the $^{16}$O+$^{184}$W and $^{19}$F+$^{181}$Ta reactions leading to the same $^{200}$Pb compound nucleus has been studied theoretically to
explain the experimental data showing more intense yield of evaporation residue in the former reaction in comparison with that in the latter reaction. The observed difference is explained by large capture cross section in the former
 and  by  increase of the quasifission  contribution  to the yield of fission-like fragments in the $^{19}$F+$^{181}$Ta reaction at large excitation energies. The probability of compound nucleus formation in the $^{16}$O+$^{184}$W reaction is larger  but  compound nuclei formed in both reactions have similar angular  momentum ranges at the same excitation energy.  The  observed decrease of evaporation residue cross section normalized to the fusion cross section in the $^{19}$F+$^{181}$Ta reaction in comparison with the one in the $^{16}$O+$^{184}$W reaction at high excitation energies is explained by the increase of hindrance in the formation of compound nucleus connected with more quick increase of the quasifission contribution in the $^{19}$F induced reaction. The spin distributions of the evaporation residue cross sections for the two reactions are also presented.
\end{abstract}

\end{frontmatter}


\section{Introduction}
\label{Introd}
The observed yield of evaporation residues (ER) in experiments is a result of
the de-excitation of a heated and rotating compound nucleus formed
in complete fusion reactions in competition against fission at heavy ion collisions.
The experimental values of its cross section are measured with the enough
 good accuracy and it is the leading light for the theoretical models to fix their parameters
or for their improvement.

In Ref. \cite{ShidlingPLB} the difference in the  excitation functions of the
ER formation in the $^{16}$O+$^{184}$W \cite{ShidlingPRC} and
$^{19}$F+$^{181}$Ta \cite{HindeNPA385}
reactions leading to the same $^{200}$Pb compound nucleus was observed
in analysis of the measured data. The
comparison of experimental results of both the
systems shows that ER cross sections and moments of the gamma multiplicity
distribution of the $^{16}$O+$^{184}$W reaction are significantly higher
than those of the $^{19}$F+$^{181}$Ta system at higher excitation energies.
Authors concluded that the reduction in the ER cross sections and moments
of the spin distribution for the $^{19}$F+$^{181}$Ta reaction is mainly due
to the suppression of fusion of higher values of the orbital angular momentum.
The present paper is devoted to the theoretical study of this observed
difference in the  excitation functions of the ER formation in the
$^{16}$O+$^{184}$W and $^{19}$F+$^{181}$Ta reactions. Our analysis
in the framework of the dinuclear system model
\cite{FazioEPJ2004,GiaEur2000,FazioPRC2005,NasirovNPA759} and advanced statistical model \cite{ArrigoPRC1992,ArrigoPRC1994,SagJPG1998} shows that
an appearance of the difference between the presented values of the ratio
of the evaporation residues to complete fusion cross sections in Ref.
\cite{ShidlingPLB} is explained by the increase of the quasifission
and fast fission contributions  to the measured fission fragments.
As well as the difference in the formation of compound nuclei and
their spin distributions are explored to explain the dependence of
the  $\sigma_{ER}/\sigma_{fus}$ ratio as a function of excitation
energy and angular momentum.

There are two main reasons causing a hindrance to the ER formation
in the reactions with massive nuclei: the quasifission and
fusion-fission processes. The effects of these binary processes prove in different stages of reaction. The angular and mass distributions of some part of their
products  can overlap. The ER formation process is often
considered as third stage of the three-stage process. The first
stage is a capture--formation of the dinuclear system (DNS) after
full momentum transfer of the relative motion of colliding nuclei
into the deformed shape, excitation energy and rotational energy.
The capture takes place if the initial energy of projectile in the
center-of-mass system is enough to overcome the interaction
barrier (Coulomb barrier + rotational energy of the entrance channel).
The study of dynamics of processes in heavy ion collisions at energies near
the Coulomb barrier  shows that complete fusion does not occur immediately in the case of the massive nuclei collisions \cite{FazioEPJ2004,Back32,VolPLB1995,AdamianPRC2003,
NasirovPRC79}. The quasifission process competes with formation of compound
nucleus (CN). This  process occurs when the dinuclear system
prefers to break down into fragments instead of to be transformed
into a fully equilibriated  CN. The number of events going to
quasifission increases drastically by increasing  the sum of the
Coulomb interaction and rotational energy in the entrance channel
\cite{GiaEur2000,FazioPRC2005,NasirovPRC79}. Another reason decreasing yield of
ER by increasing excitation energy is usual fission of a heated and rotating CN which was formed in competition with quasifission. The stability of a massive CN
decreases due to the decrease of the fission barrier by increasing
its excitation energy $E^*_{CN}$ and angular momentum $L=\ell \hbar$
\cite{ArrigoPRC1992,ArrigoPRC1994,SagJPG1998}.
In collisions with large values of the
orbital angular momentum the yield of ER decreases due to  the
 fast fission of being formed compound nucleus. The fast fission is the
inevitable decay of the fast rotating mononucleus into two
fragments without reaching the equilibrium compact shape of CN \cite{Gregoire}.
Such mononucleus is formed from the dinuclear system survived
against quasifission but it immediately decays into two fragments
 if the value of its angular momentum
is larger than $\ell_f$  at which the
fission barrier of the corresponding CN disappears. So the fast fission
process takes place at $\ell > \ell_f$. Distinct from fast fission,
the quasifission can occur at all values of $\ell$ at which the capture occurs
\cite{NasirovNPA759,NasirovPRC79}. So, the main channels decreasing the cross section  of complete fusion are quasifission and fast fission processes.
Furthermore these channels produce binary fragments which can overlap
with the ones of the fusion-fission channel and the amount  of the mixed detected fragments  depends on the mass asymmetry of entrance channel,
as well as the shell structure of the being formed reaction fragments.
Therefore,  the correct estimation of the cross section of the compound nucleus
formation in the reactions with massive nuclei is enough
difficult task for both experimentalists and theorists.
Different assumptions about the fusion process are used
in different theoretical models and they can give different
cross sections.

The experimental methods used to estimate the fusion
probability depend on the unambiguity of identification of the
complete fusion reaction products among the quasifission products. The
difficulties arise when the mass (charge) and angular
distributions of the quasifission and fusion-fission fragments
strongly overlap depending on the reaction dynamics.
As a result, the complete fusion cross sections may be overestimated
\cite{NasirovPRC79}.
We think the compared ratios of cross sections between evaporation residues and complete fusion
\begin{equation}
R=\sigma_{ER}/\sigma_{fus}.
\end{equation}
for the $^{16}$O+$^{184}$W and $^{19}$F+$^{181}$Ta reactions
\cite{ShidlingPLB} are not free from the influence of the above mentioned
ambiguity in determination of the fusion cross section $\sigma_{fus}$.  The  experimental value of $\sigma_{fus}$ reconstructed by the detected  fission fragments and evaporation residues  can be contributed by the following terms
\begin{equation}
\sigma_{\rm fus}^{(\rm exp)}=\sigma_{\rm ff}^{(\rm exp)}+\sigma_{\rm ER}^{(\rm exp)}+
\sigma_{\rm qf}^{(\rm exp)}+\sigma_{\rm fast\,fis}^{(\rm exp)},
\end{equation}
where $\sigma_{\rm ff}^{(\rm exp)}$, $\sigma_{\rm qf}^{(\rm exp)}$ and
$\sigma_{\rm fast\,fis}^{(\rm exp)}$  are the contributions of fusion-fission, quasifission and fast-fission processes, respectively, and $\sigma_{\rm ER}^{(\rm exp)}$ is the ER contribution. According to the statement of authors
of Ref. \cite{ShidlingPLB} the complete fusion cross sections are obtained by adding fission cross section \cite{Sinha} to the measured data of the evaporation residue cross sections.
In Ref. \cite{Forster} the complete fusion cross section is derived from a statistical model where only neutron evaporation and fission are included.
We think that the used fission data from Ref. \cite{Sinha} contain quasifission and in some cases also fast fission contributions which appear as hindrance to the complete fusion. This argument is confirmed by our results obtained in the framework of the dinuclear system model. The total ER and fusion-fission excitation functions are  calculated by us in the framework of the advanced statistical model \cite{ArrigoPRC1992,ArrigoPRC1994,SagJPG1998}.

The evaporation residue cross section normalized to fusion cross section which was analyzed in Ref. \cite{ShidlingPLB} can be presented as follows
\begin{equation}
\label{ratioexp3}
R^{(\rm exp)}=
\sigma_{\rm ER}^{(\rm exp)}/(\sigma_{\rm ff}^{(\rm exp)}+\sigma_{\rm ER}^{(\rm exp)}+
\sigma_{\rm qf}^{(\rm exp)}+\sigma_{\rm fast\,fis}^{(\rm exp)}).
\end{equation}
The  true values of $R=\sigma_{\rm ER}^{(\rm exp)}/\sigma_{\rm fus}$ can be found from (\ref{ratioexp3}):
\begin{equation}
\label{ratio}
R=R^{(\rm exp)}(1+(\sigma_{\rm qf}^{(\rm exp)}+
\sigma_{\rm fast\,fis}^{(\rm exp)})/\sigma_{\rm fus}).
\end{equation}
where $\sigma_{\rm fus}$ is the pure fusion cross section ($\sigma_{\rm ER}+\sigma_{\rm ff}$).
The equation (\ref{ratio}) means that a presence of the quasifission and fast fission contributing to the measured data of fusion cross section leads to decrease the ratio between the evaporation residue and fusion cross sections.

\section{Study of difference in the evaporation residue
formation in the $^{16}$O+$^{184}$W and $^{19}$F+$^{181}$Ta reactions}
\label{Study}

Authors of Ref. \cite{ShidlingPLB} studied the role of the entrance channel
in formation of the evaporation residue obtained by comparison of the
ratio $R^{(\rm exp)}$ (\ref{ratioexp3}) of the measured evaporation residue
cross section $\sigma_{ER}$ to the fusion cross section $\sigma_{fus}$  in the $^{16}$O+$^{184}$W and $^{19}$F+$^{181}$Ta reactions as a function
of the excitation energy  $E^*_{\rm CN}$ and angular
momentum of compound nucleus. The ratio $R^{(\rm exp)}$ for both
reactions is approximately equal up to  excitation energy $E^*_{\rm CN}$ of about 67 MeV and at larger excitation energies the values  of $R^{(\rm exp)}$ corresponding to the $^{16}$O+$^{184}$W reaction become larger  than that for the
$^{19}$F+$^{181}$Ta reaction (see in forward Fig. \ref{fig3} on the left axis). In Ref. \cite{ShidlingPLB}, the observed
 deficiency in the ER cross sections for the $^{19}$F+$^{181}$Ta
 system is explained  by the suppression of partial ER cross sections at higher spin values.  So, authors of Ref. \cite{ShidlingPLB} stressed the importance
of spin distribution measurements.

Our calculations show that the difference between the
ratios $R^{(\rm exp)}$ for the $^{16}$O+$^{184}$W and $^{19}$F+$^{181}$Ta reactions
 at large $E^*_{\rm CN}$  energies appears due to the more large contributions
 of quasifission and fast
 fission into the measured fission fragments in the latter reaction.
To understand the appearance of this difference, the fusion
cross section has to be analyzed because it is a quantity
causing ambiguity in $R^{(\rm exp)}$ connecting with the identification of the
true fusion-fission
products. In  Ref. \cite{ShidlingPLB} the total fusion cross
sections were obtained by addition of the fission \cite{Sinha} and
evaporation residue cross sections \cite{Forster}.
\begin{figure}[h]
\vspace*{0.2cm}
\begin{center}
\resizebox{0.80\textwidth}{!}{\includegraphics{{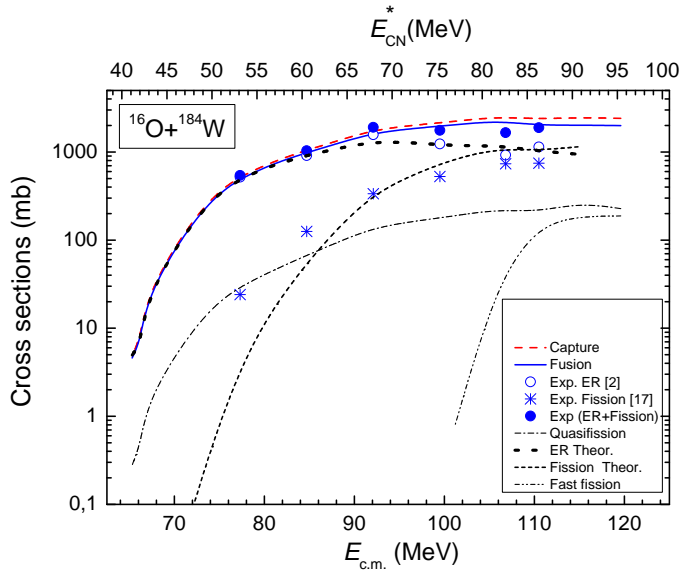}}}
\vspace*{-3.75 cm} \caption{\label{fig1} Comparison of the
experimental values of the fusion (full circles) \cite{ShidlingPRC},
evaporation residues (open circles) \cite{ShidlingPRC} and
fission excitation function (stars) \cite{Forster} for the $^{16}$O+$^{184}$W reaction with the
theoretical results obtained by the
dinuclear system model for the capture (dashed line), complete
fusion (solid line), evaporation residues (dotted line), quasifission (dot-dashed line),
fusion-fission (short dashed line), and fast fission (dot-dot-dashed line).}
\vspace*{-0.2cm}
\end{center}
\end{figure}
Therefore, in this work the difference in the formation of compound nuclei and
their spin distributions in the reactions under discussion are explored as functions
of the excitation energy and angular momentum of compound nucleus.
We estimated these contributions and the values of the fusion, fusion-fission,
ER, quasifission, and fast fission cross sections. Our results are compared with the
available experimental values for the $^{16}$O+$^{184}$W \cite{ShidlingPRC} and
$^{19}$F+$^{181}$Ta \cite{HindeNPA385} systems as reported in Figs.
\ref{fig1} and \ref{fig2}, respectively.
\begin{figure}[h]
\vspace*{1.15cm}
\begin{center}
\resizebox{0.80\textwidth}{!}{\includegraphics{{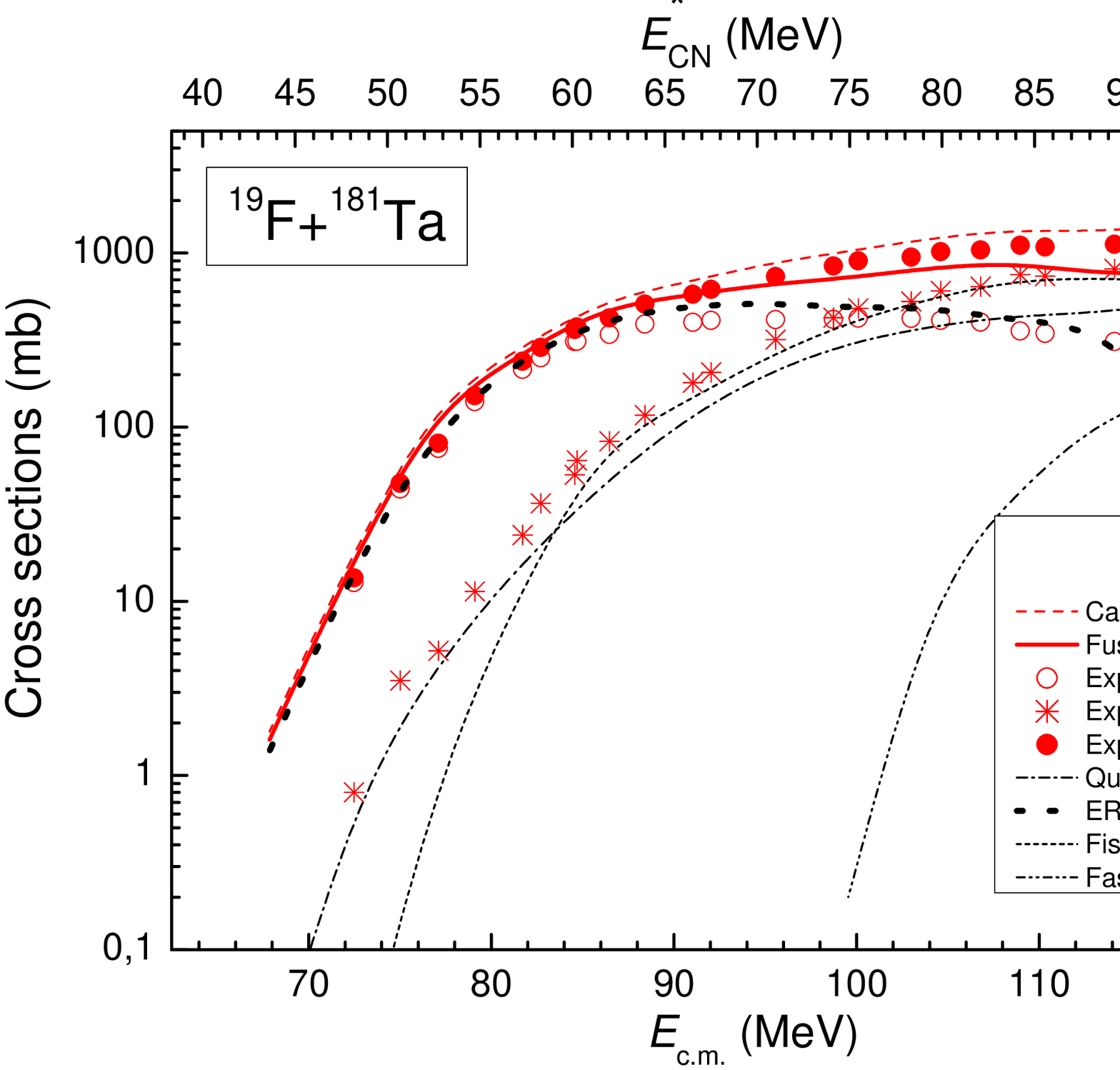}}}
\vspace*{-3.75 cm} \caption{\label{fig2} As in Fig. \ref{fig1} but for the $^{19}$F+$^{181}$Ta reaction.}
\vspace*{-0.4 cm}
\end{center}
\end{figure}
The measured total value of the ER cross section for the $^{16}$O+$^{184}$W  reaction is much larger than that for the $^{19}$F+$^{181}$Ta system. This means that the complete fusion cross section for the former reaction is larger too.  The significant difference between the fusion probabilities for the reactions under discussion is caused by the large capture probability for the $^{16}$O+$^{184}$W reaction because the potential well of the nucleus-nucleus interaction for the more asymmetric system is wider and deeper. Therefore, the measured cross sections  of the fission-like and ER fragment yields for the $^{16}$O+$^{184}$W reaction are larger than the ones for the $^{19}$F+$^{181}$Ta reaction (see Figs. \ref{fig1} and \ref{fig2}).  Another reason causing a hindrance at formation of compound nucleus in the $^{19}$F+$^{181}$Ta reaction at large energies is the increasing contribution of the quasifission process. The theoretical values of the quasifission contribution are presented in Fig. \ref{fig2} by the dot-dashed line while the fast fission is represented by dot-dot-dashed line. Theoretical values of the fusion-fission cross section are shown by short-dashed line.
From our results we can conclude that at low energies the fission-like fragments are mainly quasifission products. At beam energies corresponding to the excitation energy $E^*_{\rm CN}$ of about 62 MeV  the yield of the fusion-fission and quasifission fragments become comparable and at higher energies the fission cross section overcomes the one of quasifission. The fast fission fragments contribute not so strongly and their maximum value is about 15\% of the observed yield of fission-like fragments only
at large excitation energies $E^*_{\rm CN}>85$ MeV.
\begin{figure}
\vspace*{0.4cm}
\begin{center}
\resizebox{0.90\textwidth}{!}{\includegraphics{{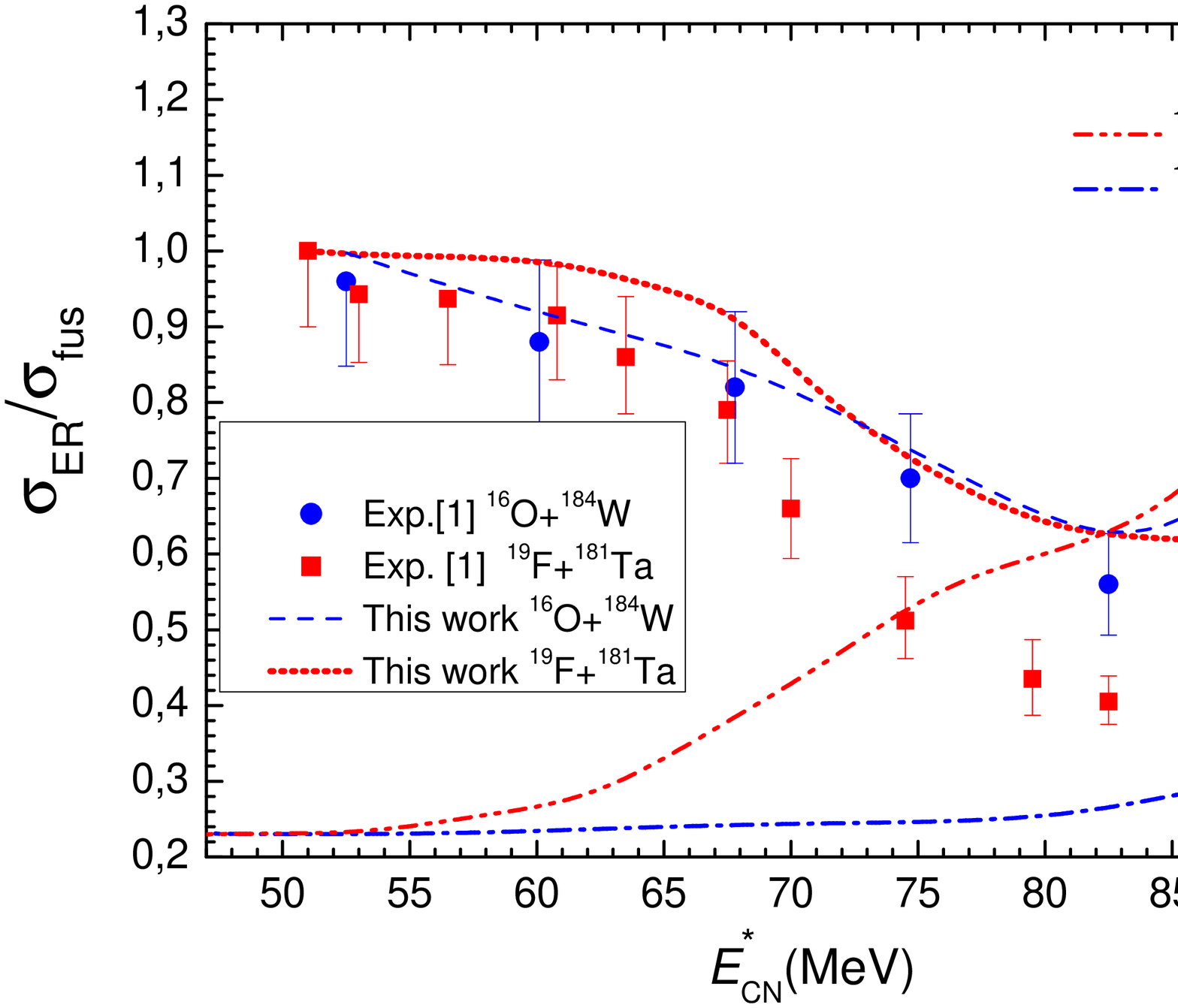}}}
\vspace*{-4.35 cm} \caption{\label{fig3} Comparison of the experimental values of the evaporation
residue cross sections (normalized with respect to the fusion cross sections)
for the $^{16}$O+$^{184}$W (solid circles) and $^{19}$F+$^{181}$Ta systems
(solid squares) \cite{ShidlingPLB} with the corresponding theoretical results (dashed and dotted lines, respectively) as a function of the excitation
energy $E^*_{\rm CN}$ of compound nucleus (left axis). Theoretical results
of the sum of the quasifission and fast fission cross sections (normalized with respect of the fusion cross sections) for the $^{16}$O+$^{184}$W (dot dashed line) and $^{19}$F+$^{181}$Ta (dot-dot dashed line) systems are presented versus $E^*_{\rm CN}$ and compared on the right axis.}
\vspace*{-0.6 cm}
\end{center}
\end{figure}

\begin{figure}[h]
\vspace*{0.25cm}
\begin{center}
\resizebox{1.00\textwidth}{!}{\includegraphics{{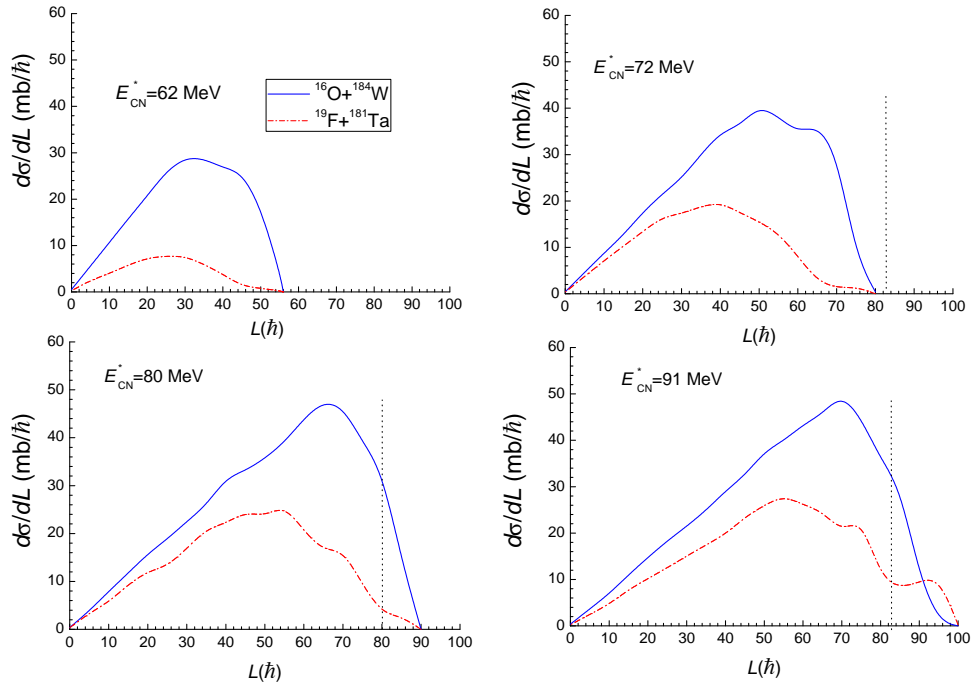}}}
\vspace*{-4.85 cm} \caption{\label{fig4} Partial fusion cross sections as a function
of the angular momentum for the $^{16}$O+$^{184}$W (solid line) and $^{19}$F+$^{181}$Ta
(dashed line) reactions at the excitation energies $E^*_{\rm CN}=$62, 72, 80, and 91 MeV.
The vertical dashed line at $L_{\rm f}=80 \hbar$ separates complete fusion and
fast fission reactions (about $L_{\rm f}$ see text).}
\end{center}
\end{figure}

In  Fig.  \ref{fig3} the results of evaporation residue cross sections (normalized with respect to the fusion cross sections)
for the $^{16}$O+$^{184}$W (solid circles) and $^{19}$F+$^{181}$Ta
(solid squares) systems \cite{ShidlingPLB} are compared with the corresponding theoretical
values (dashed and dotted lines for the two reactions, respectively)
for the corrected ratio $R$ (left axis) as a function of the excitation
energy $E^*_{\rm CN}$ of compound nucleus. The theoretical curves of $R$, obtained by formula (\ref{ratio}) with the aim of excluding the quasifission and fast fission contributions in the experimental fusion cross section and determining the true fusion cross sections, are higher than the experimental points  used in Ref. \cite{ShidlingPLB}. For the $^{16}$O+$^{184}$W  reaction the curve $R$ (dashed line) is  a bit higher than the experimental points (almost within the error bars) because the effect of the quasifission and fast fission contributions on the used experimental fusion cross section is small (see the trend of dash-dotted line, and read the sum of the quasifission and fast fission cross sections normalized with respect to the fusion cross section on the right axis of Fig.  \ref{fig3}).  For the $^{19}$F+$^{181}$Ta reaction the curve of $R$ (dotted line) is close to the results of $R$ for the $^{16}$O+$^{184}$W  reaction  in the  $E^*_{\rm CN}$=50--67 MeV  energy range where the experimental points for the two reactions are similar within the error bars.
At energies higher than  67 MeV the calculated results of  $R$  for the two reactions are  in complete agreement each with other while the experimental points for the $^{19}$F+$^{181}$Ta reaction deviated from the ones for the  $^{16}$O+$^{184}$W  reaction.
 The closest of values of the evaporation residue cross sections normalized with respect to the fusion cross sections for the reactions under discussion means that survival probability of the compound nucleus formed in both reactions has the same dependence on the excitation energy $E^*_{\rm CN}$. The reason of the deviation between experimental values
of $R^{\rm (exp)}$ for the $^{19}$F+$^{181}$Ta reaction (solid squares) from the ones of the $^{16}$O+$^{184}$W reaction (solid circles)  at large excitation energies $E^*_{\rm CN}> 67$ MeV is explained by the increasing contribution of quasifission into measured fission cross sections for the former reaction (see Fig.\ref{fig2}). The growing contribution of quasifission in the $^{19}$F+$^{181}$Ta reaction is also clearly shown by the trend of the $\sigma_{(qf + fast\, fission)}/\sigma_{fus}$ ratio versus $E^*_{\rm CN}$ presented  on the right axis of Fig.  \ref{fig3}. This  effect of the reaction dynamics in the entrance channel on the reaction products appears  in the $^{19}$F+$^{181}$Ta reaction at higher energies, while it is small in the $^{16}$O+$^{184}$W  reaction.
 It means that the lower cross section of the evaporation residue for the $^{19}$F+$^{181}$Ta reaction is connected with the capture and complete fusion stages.  This conclusion is supported by the comparison of the angular momentum distribution of compound nuclei formed in these reactions. The results presented in Fig. \ref{fig4} for the excitation energies $E^*_{\rm CN}=$62, 72, 80, and 91 MeV show that the spin distributions of compound nuclei differ mainly by the probability but not by the values of angular momentum ranges. This means that the yields of compound nuclei formed in both reactions under discussion are different but they have similar range of the angular momentum $L$. The vertical dotted lines at $L_{f}=80\hbar$ separates the complete fusion and fast fission regions of angular momentum.

\begin{figure}
\vspace*{1.5cm}
\begin{center}
\resizebox{0.90\textwidth}{!}{\includegraphics{{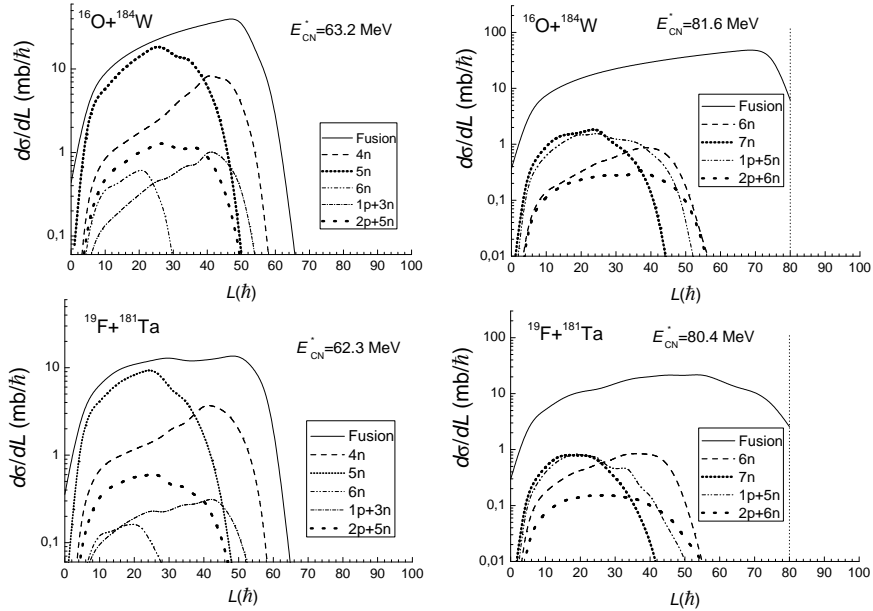}}}
\vspace*{-4.35 cm} \caption{\label{fig5}
Spin distribution of evaporation
residue cross sections as a function of the angular momentum $L$. The upper part is for  the $^{16}$O+$^{184}$W reaction, the lower part is for $^{19}$F+$^{181}$Ta reaction, both at two $E^*_{\rm CN}$ energies: about 62-63 MeV and 80-81 MeV.}
\end{center}
\end{figure}

Moreover, in Fig. \ref{fig5} we report  the spin distributions of evaporation residue cross sections calculated by us for  $E^*_{\rm CN}$  values about 62-63  and 80-81 MeV as an example for the two reactions.
This figure shows that in the both cases of the considered $E^*_{\rm CN}$  values (the one at energy lower than  $E^*_{\rm CN}=$67 MeV, the other higher than 67 MeV) the corresponding evaporation residues  (for example, the residues after 4n, 5n, 6n, 1p+3n, 2p+5n emissions) obtained in the two reactions  cover the same angular momentum range.

In conclusion,  we distinguish two points: i) the apparent different behavior of the experimental values of the   $\sigma_{\rm ER}/\sigma_{\rm fus}$  ratio  is not due to the different decay dynamic  of  compound nuclei formed in the $^{16}$O+$^{184}$W and $^{19}$F+$^{181}$Ta reactions but it is due to the  different quasifission contributions which can be considered as hindrance to complete fusion in the entrance channel; ii)  the different yields of ER's and fusion-fission fragments for the two reactions are caused by different capture cross sections formed at the first stage of the reacting nuclei.  Different quasifission contributions causing in the $^{19}$F+$^{181}$Ta reaction a relevant hindrance to fusion and consequently in the ER and fusion-fission product formations. The quasifission fragments contaminate the detected fission fragments and the determination of the fusion cross section.

In fact, if we exclude in the detected experimental fission fragments the
quasifission and fast fission contributions we obtain a good agreement
in the calculations of the $\sigma_{\rm ER}/\sigma_{\rm fus}$
ratios for the two studied reactions.  Our results of the spin distributions
of compound nuclei and evaporation residues  demonstrate the same de-excitation
dynamics of the formed compound nuclei in the two reactions.

\textbf{Acknowledgments}

A. K. Nasirov is grateful to the Istituto Nazionale  di Fisica Nucleare and
Department of Physics of the University of Messina for the support received in the collaboration between the Dubna and Messina groups and he thanks the Russian Foundation for Basic Research for the financial support at performance of this work.



\end{document}